# The Experiment and results of Laser Ranging to Space Debris[*]

Zhang Zhongping[1,2], Yang Fumin[1], Zhang Haifeng[1], Wu Zhibo[1,2], Chen Juping[1,2], Li Pu[1] and Meng Wendong[1]

[1] Shanghai Astronomical Observatory, Chinese Academy of Sciences, Shanghai 200030, China; zzp@shao.ac.cn

[2] Key Laboratory of Space Object and Debris Observation, Chinese Academy of Sciences, Nanjing 210008

**Abstract**  Space debris is a major problem for all space-active nations. Adopting high precision measuring techniques will help to produce the reliable and accurate catalogue for space debris and collision avoidance. Laser Ranging is a kind of real-time measuring technology with high precision for space debris observation. The first experiment of laser ranging to the space debris in China was performed at the Shanghai Observatory in July 2008 at the ranging precision of about 60-80cm. The experiment results show that the return signals from the targets with the range of 900 km were quite strong with the power of 40W (2J@20Hz), 10ns pulse width laser at 532nm wavelength. The performances of preliminary laser ranging system and the observed results in 2008 and 2010 are introduced in the paper.

## 1 INTRODUCTION

China has launched many spacecrafts into space and had also produced some space debris since 1970s. China is also one of the members of IADC (Inter-Agency Space Debris Coordination Committee). It is necessary for China to pay great attention to reduce damages from space debris in cooperation with international community and develop kinds of high precision measuring techniques for the reliable and accurate catalogue of space debris. Laser Ranging is a kind of real-time measuring technology with high precision for space debris observation. In Oct. 2002, Dr. Ben Greene presented a report named "Laser Tracking of Space Debris" in the 13$^{th}$ International Laser Ranging Workshop and announced that they could track the space debris with the size of 15cm, 1250km distance by using the aperture of 76cm telescope and high power laser (Ben Greene, 2002).

Recent years only several countries have been doing the research on space debris laser ranging technology. The project of laser ranging to space debris at Shanghai Astronomical Observatory in China is supported by the Chinese Space Agency. A preliminary experimental laser ranging system for space debris at Shanghai Satellite Laser Ranging (SLR) station with the aperture of 60 cm telescope was set up in 2006. The major goal of the system is to develop the key techniques for laser ranging to space debris. After some testing and upgrading, we have obtained some laser returns from the discarded Soviet rocket (Catalog Number 17912) and US

[*]Supported by the Instrument Developing Project of the Chinese Academy of Sciences, Grant No. 2920100701

rocket (Catalog Number 30778) in July 2008 and the range is more than 900 km with the power of 40W laser made by a domestic institute. After that, the experimental system was improved in 2010 and several passes of laser ranging to space debris were obtained at the precision of 50-70 cm. The maximized range of the targets is up to 1200 km with 10W power laser imported from US.

The preliminary experimental system, measuring results and data analysis are given in the paper.

## 2 Performances of the laser ranging system

The experiment was carried out at the SLR station of Shanghai Astronomical Observatory, Chinese Academy of Sciences. Fig.1 shows the structure of space debris laser ranging system, including orbit prediction, control system, high power laser, laser beam transmitting system, telescope mount tracking system, high precise timing system, return detection and receiving system.

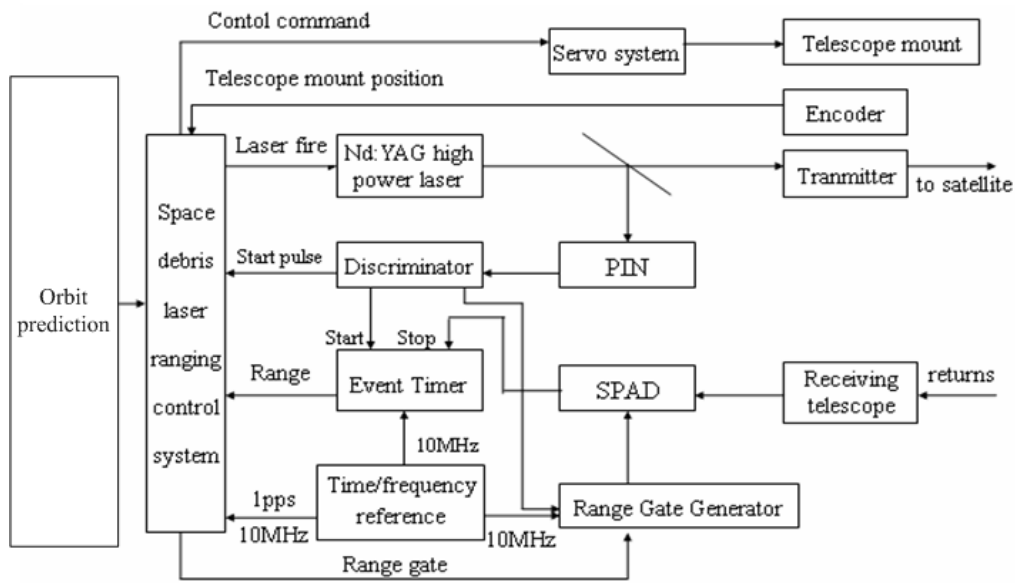

Fig.1 The structure of space debris laser ranging system

Fig.2 shows the observation house, the tracking telescope and the electronics room. The satellites equipped with retro-reflectors are routinely measured at the station. The aperture of the receiving telescope and transmitter is 60 and 21cm respectively. The mount is Alt-Azimuth type, directly driven with motors. The pointing accuracy of the telescope after star calibration is about 5 arcsec.

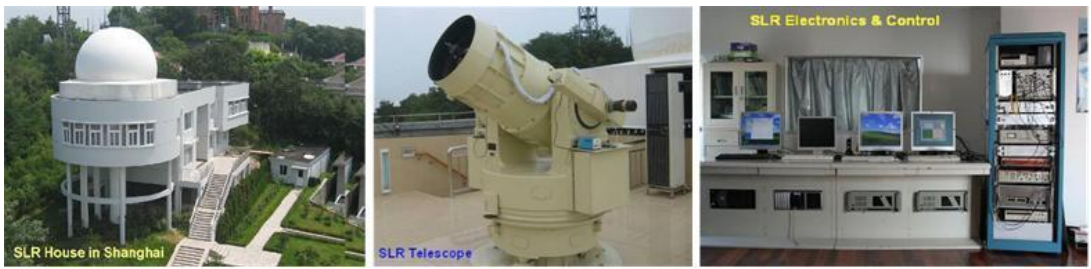

Fig.2 Shanghai SLR station, telescope and electronics control room

One of the key instruments, high power Nd:YAG laser used for the experiment was built by the North China Research Institute of Electro-Optics (NCRIEO) in Beijing. The parameters of the laser are as follows: 2J per pulse, 10nsec pulse width, 0.6 mrad divergence, 20Hz repetition rate, 40W mean power at 532nm wavelength. Fig.3 shows the block diagram of the 40W Nd:YAG laser and Fig. 4 shows the photo, inner view and laser beam through the transmitter to the sky.

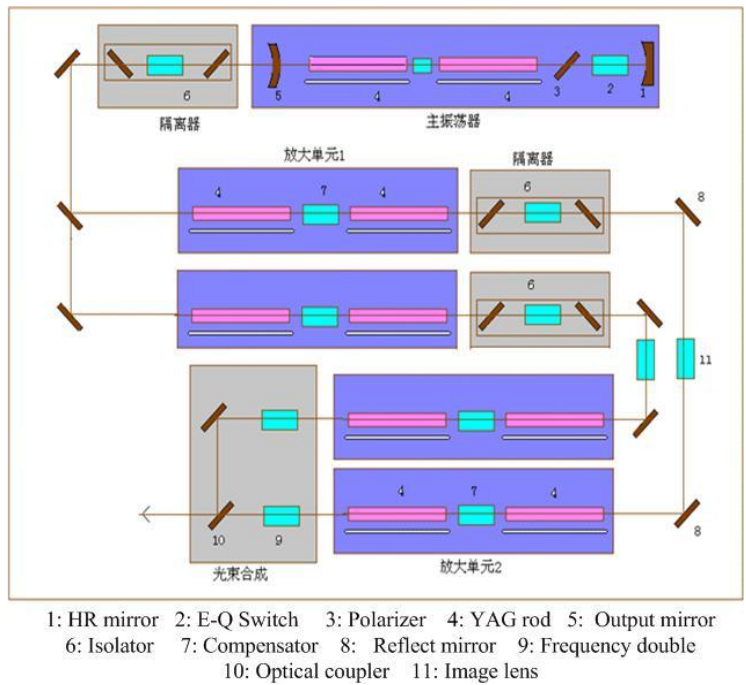

Fig.3 Diagram of the 40W Nd:YAG Laser (2J in 532nm, 20Hz)

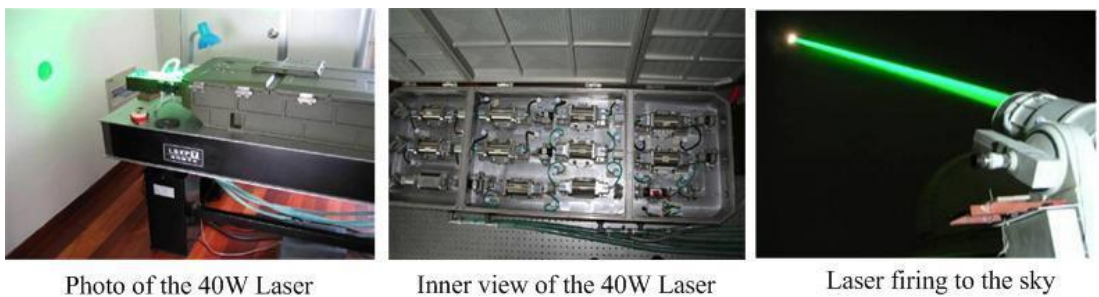

Fig.4 Photo, Inner view and laser beam of the 40W Laser

There are ten Nd:YAG rods in total in the laser system. The output from the oscillator with two laser rods inside is divided into two beams, and then goes to the amplifier units 1 and 2 respectively. The outputs from the two sets of amplifier unit pass the frequency doublers, and then combine into one beam for ranging.

The detector and the time interval instrument are adopted the ones as the routine SLR operation. Due to space debris of irregular shape and laser pulse-width, the measurement of space debris is lower than the routine SLR. The single photon avalanche diode detector with single photon sensitivity and 30psec timing precision was provided by the Czech Technical University. The event timer Model A032-ET for time interval measurement with 10 psec timing precision was made by the Riga University, Latvia(Artyukh, 2001, 2008).

## 3 The analysis of return detected by laser radar equation

The returned signal strength expected from a 2-meter diameter target located 800 km away can be roughly estimated by the following equation (Degnan, 1993):

$$n_0 = \frac{\lambda \eta_q}{hc} \times \frac{E_t A_r \rho S \cos\theta}{\pi \theta_t^2 R^4} \times T^2 \times K_t \times K_r \times \alpha$$

Where:

- $n_0$ : Average number of photoelectrons received by detector
- $\lambda$ : Wavelength of Laser, 532nm
- $\eta_q$ : Quantum efficiency of the detector, 0.2
- $h$ : Planck constant, $6.6260693 \times 10^{-34}$ J s
- $c$ : Light speed, 299792458 m/s
- $E_t$ : Energy of laser pulse, 2J
- $A_r$ : Effective area of receive telescope, $0.245 \mathrm{m}^2$
- $\rho$ : Reflectivity of the targets surface, 0.16
- $S$ : Effective reflective area of target. Equivalent radius is 1 m, $S = \pi r^2$
- $\cos\theta$ : Suppose the targets are spherical, $\cos\theta = 1$
- $\theta_t$ : Divergence of laser beam from telescope, 12 arcsec
- $R$ : Range of the targets, 800km
- $T$ : Atmospheric transmission, $T = 0.6$ at elevation 30 degree
- $K_t$ : Efficiency of transmitting optics, 0.6
- $K_r$ : Efficiency of receiving optics, 0.6
- $\alpha$ : Attenuation factor mainly caused from atmospheric effect, 13dB

We have

$$n_0 = 0.13 \text{ (Photoelectrons)}$$

The probability of detection by the C-SPAD detector with a single photon sensitivity can be estimated by

$$P = 1 - e^{-n_0} = 1 - e^{-0.13} = 0.12$$

So we might get 12 returns in the observation time interval of each 5 second with the 20Hz laser theoretically.

## 4 Observational Results from the space debris at Shanghai Observatory

After the installation of the 40W power laser, we built the control and ranging interfaces and software for the experiment. In July of 2008, we firstly obtained some laser returns from the space debris which were the discarded Soviet and US rockets with the catalog number 17912 (639×611 km) and 30778 (541×499 km) respectively. The range residuals (O-C) via the elapse time for three passes are shown in Fig.5 and ranging precision is from 60 to 80cm. In Fig.5, the horizontal axis is elapse time from beginning of tracking and the vertical axis is range residuals O-C between the observed and predicted. The central points in line are the laser returns, the other points are noises from the detector and background.

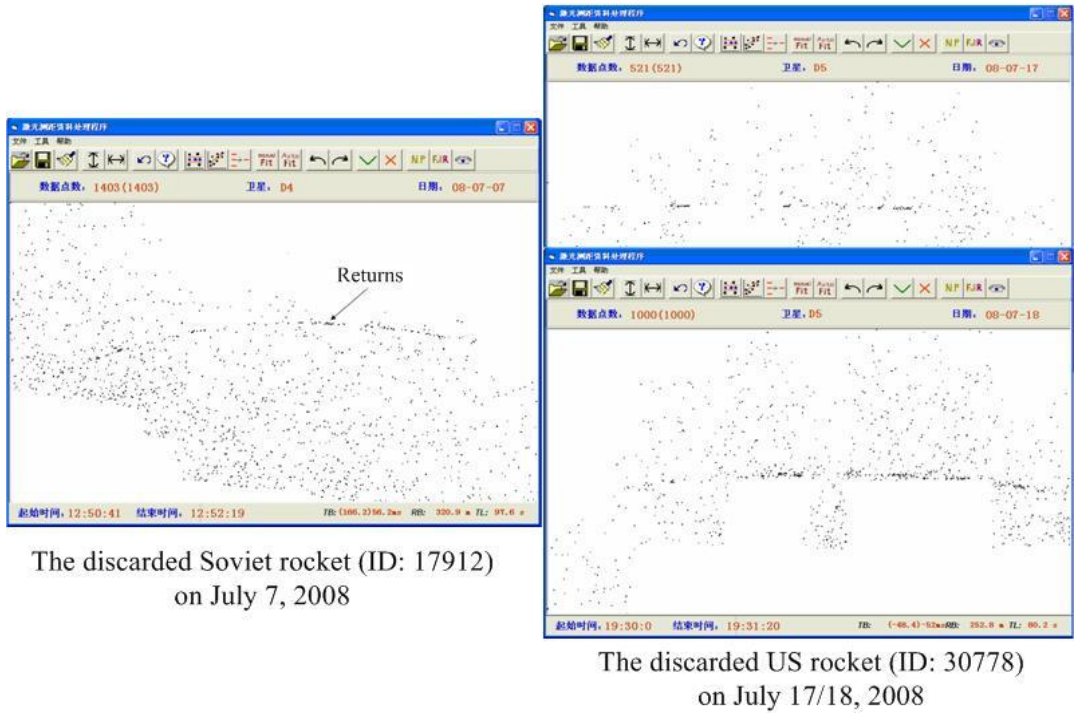

Fig.5 Laser returns from the space debris

Fig.6 shows the range variations for each pass, the horizontal axis is seconds of the day and the vertical axis is range of the space debris. The maximum range obtained in the measurement was 936 km.

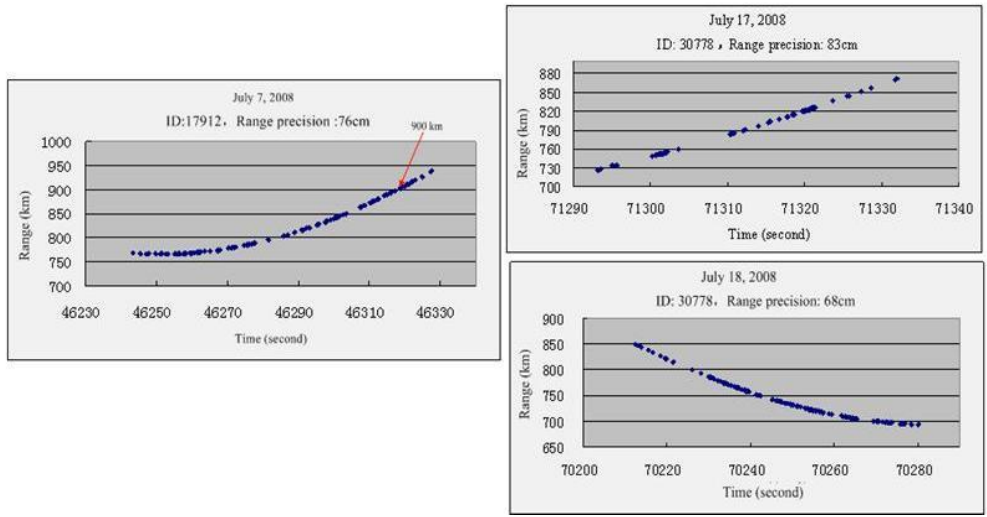

Fig.6 Obtained ranging data of space debris

Fig.7 shows the statistics of the laser returns in 5 second bins from the ID 17912 on July 7, 2008. The horizontal axis is time binning in 5 seconds and the vertical axis is the number of returns obtained. Laser return rate is about 7%. It is shown that about 10 to 14 returns in several 5-second interval were obtained when the telescope's tracking was good, and it roughly coincides with the theoretic estimation of the returned signal strength.

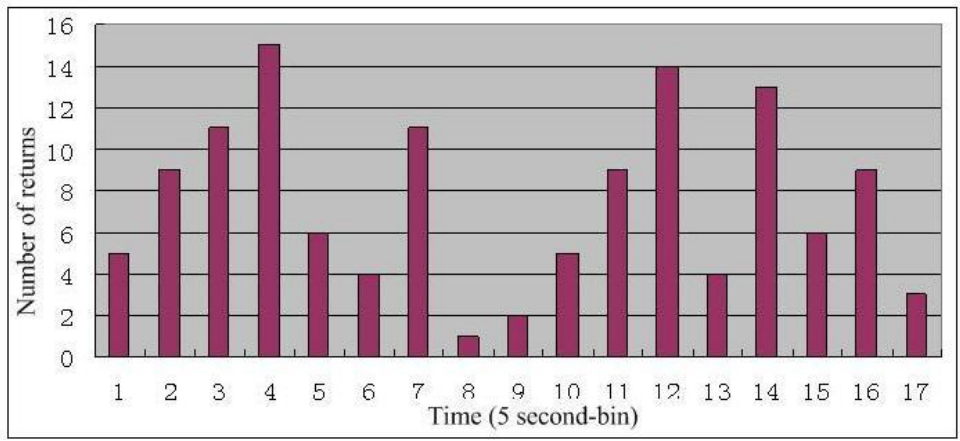

Fig.7 Statistics of laser returns (5-second time interval bin) on July 7, 2008

In 2010, Shanghai SLR station updated the experimental system for space debris laser ranging, including adopting better stable high power laser (1J@10Hz, output 10W at 532nm wavelength), advancing the ability of servo tracking system, multi step range gate adjusting automatically and applying Two Line Elements (TLE) predicted orbit with the accuracy of less than 1 km. The capacity of experimental system of laser ranging to space debris was improved obviously. Fig.8 shows the servo tracking performances of telescope mount and the tracking precision is less than 1 arcsec.

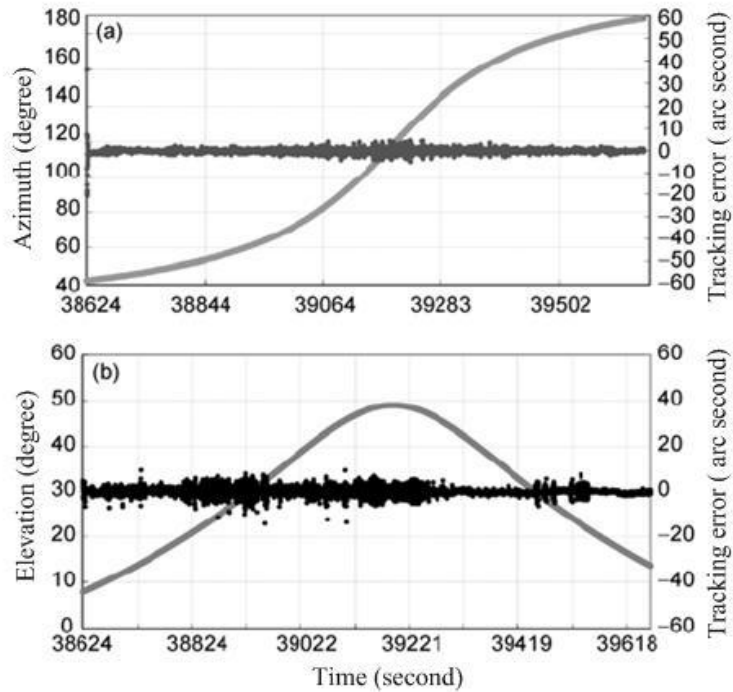

Fig.8 Tracking performance of telescope mount

Fig.9 shows some measuring results obtained by the 10W power laser. The range measured is from 800 to1200 km, ranging precision from 50 to 80cm. The most returns are up to more than 150 points within 66 seconds and the return rate is about 23%. It can be seen that the ability of ranging system by using 10W laser is improved compared to the one using 40W laser.

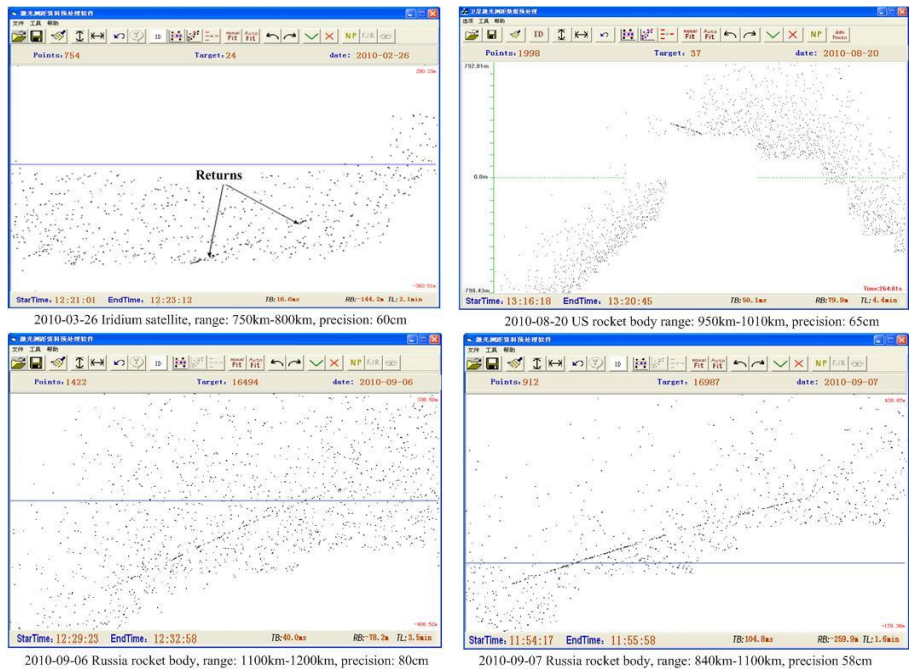

Fig.9 Some measuring results by 10W power laser in 2010

The above measuring results show that the experimental system for space debris laser ranging established by Shanghai Astronomical Observatory has realized its preliminary goal and had the ability to track the large-scale, low orbit space debris by laser technology. Although some improved methods were adopted, the difficulty of laser ranging to space debris is quite hard, mainly because of the power of the laser, the accuracy of predicted orbit, the uncertainty of reflective characteristic on the surface and the size of space debris. So some advanced methods should be developed to further research the technology of space debris laser ranging.

## 5 CONCLUSIONS

The experiment of laser ranging to space debris in China was successfully performed at the Shanghai Observatory in July 2008 and several passes of laser ranging to space debris were obtained. It is shown from the experiment that the return signals from the targets with the range of 900 km were quite strong. This experiment verified the high accuracy method of the real-time laser-determined orbital elements and explored a wide range of Chinese satellite laser ranging technology. Through improving the preliminary laser ranging system, more passes and farther space debris were observed. However, due to many influencing factors, the measuring success rate is very low, further experiment research for laser ranging to space debris should be going on in order to measure farther and smaller targets and increasing the ability of routinely space debris laser tracking.

**Acknowledgement** The research is supported by the Chinese Space Agency. We would like to thank Dr.Josef Kolbl of Deggendorf University of Applied Sciences, Germany for his comments and suggestions for the estimate of return signal strength.